\documentclass[aps,floatfix,nofootinbib,showpacs,twocolumn,superscriptaddress]{revtex4}

\usepackage{graphicx}
\usepackage{dcolumn}
\usepackage{amsmath}
\usepackage{longtable}

\begin{document}


\title{Chiral susceptibility and the scalar Ward identity}

\author{Lei Chang（常雷）}
\affiliation{Institute of Applied Physics and Computational Mathematics, Beijing 100094, China}

\author{Yu-xin Liu（刘玉鑫）}
\email[Corresponding author: ]{yxliu@pku.edu.cn}
\affiliation{Department of Physics and the State Key Laboratory of
Nuclear Physics and Technology, Peking University, Beijing 100871,
China} 
\affiliation{Center of Theoretical Nuclear Physics, National
Laboratory of Heavy Ion Accelerator, Lanzhou 730000, China}

\author{Craig D. Roberts}
\email[Corresponding author: ]{cdroberts@anl.gov}
\affiliation{Physics Division, Argonne National Laboratory, Argonne,
Illinois 60439, USA}
\affiliation{School of Physics, The University of New South Wales, Sydney NSW 2052, Australia}

\author{Yuan-mei Shi（石远美）}
\affiliation{Department of Physics, Nanjing University, Nanjing
210093, China}

\author{Wei-min Sun（孙为民）}
\affiliation{Department of Physics, Nanjing University, Nanjing
210093, China} 
\affiliation{Joint Center for Particle, Nuclear
Physics and Cosmology, Nanjing 210093, China}

\author{Hong-shi Zong（宗红石）}
\affiliation{Department of Physics, Nanjing University, Nanjing
210093, China} \affiliation{Joint Center for Particle, Nuclear
Physics and Cosmology, Nanjing 210093, China}

\date{\today}

\begin{abstract}
The chiral susceptibility is given by the scalar vacuum polarisation at zero total momentum.  This follows directly from the expression for the vacuum quark condensate so long as a nonperturbative symmetry preserving truncation scheme is employed.
For QCD in-vacuum the susceptibility can rigorously be defined via a Pauli-Villars regularisation procedure.  
Owing to the scalar Ward identity, irrespective of the form or \emph{Ansatz} for the kernel of the gap equation, the consistent scalar vertex at zero total momentum can automatically be obtained and hence the consistent susceptibility.  
This enables calculation of the chiral susceptibility for markedly different vertex \emph{Ans\"atze}.  For the two cases considered, the results were consistent and the minor quantitative differences easily understood.  
The susceptibility can be used to demarcate the domain of coupling strength within a theory upon which chiral symmetry is dynamically broken.  Degenerate massless scalar and pseudoscalar bound-states appear at the critical coupling for dynamical chiral symmetry breaking.  
\end{abstract}


\maketitle

\section{Introduction}
The analysis of colour-singlet current-current correlators 
\begin{equation}
\langle \bar q(x) \Gamma q(x) \, q(0) \Gamma q(0)\rangle ,
\end{equation}
where $\Gamma$ is a Dirac matrix, or, equivalently, of the associated vacuum polarisations, plays an important role in QCD because these quantities are directly related to observables.  The vector vacuum polarisation, e.g., couples to real and virtual photons.  It is thus basic to the analysis and understanding of $e^+ e^- \to\,$hadrons and a determination of the strong running coupling on the perturbative domain \cite{Gonsalves:2008rd,Krein:1990sf}.  Nonperturbative information is also available.  For example, the mass of a hadron can be estimated in lattice-regularised QCD by analysing the large Euclidean-time behaviour of a carefully chosen correlator \cite{Morningstar:2005pv,Bhagwat:2007rj}.  Correlators are also amenable to analysis via the operator product expansion and are therefore fundamental in the application of QCD sum rules \cite{Colangelo:2000dp}. 

A given vacuum polarisation is a function of the total momentum, $P$, which is conjugate to the spacetime separation between the two currents.  Its value at $P=0$ yields a vacuum susceptibility; namely, a measure of the response of the theory's ground state to a fluctuation in some external parameter.  Herein we consider the chiral susceptibility, which in the chiral limit measures the response of the ground state to a fluctuation in the current-quark mass.  In the theory of phase transitions this is analogous to considering the response of a magnetisation to an infinitesimal external magnetic field.  It is fundamental to understanding the phases that may be realised in a given theory.  

In QCD the quark-parton acquires a momentum-dependent mass function, which at infrared momenta is $\sim 100$-times larger than the current-quark mass.  The Dyson-Schwinger equations (DSEs) \cite{Roberts:1994dr,Roberts:2007jh}
explain that this effect owes primarily to a dense cloud of gluons that clothes a low-momentum quark \cite{Bhagwat:2007vx,Roberts:2007ji}.  This marked momentum-dependence of the dressed-quark mass function is one manifestation of dynamical chiral symmetry breaking (DCSB), which is the single most important mass generating mechanism for light-quark hadrons; e.g., it is responsible for roughly 98\% of a proton's mass.  

The behaviour of the chiral susceptibility can be used to determine under which conditions chiral symmetry is dynamically broken; e.g., to demarcate the domain of coupling strengths upon which the phenomenon occurs.  It can also be employed to analyse the nonzero temperature behaviour of QCD, and to determine the critical temperature for chiral symmetry restoration and the critical exponents associated with that transition; e.g., Refs.\,\cite{Blaschke:1998mp,Holl:1998qs,Aoki:2006we,Fu:2007xc}.  

Our aim herein is to explain the essence and uses of the chiral susceptibility in QCD.  In this connection it is important to note that when working with a vacuum polarisation or susceptibility the question of an appropriate regularisation scheme arises \cite{Zong:2005ie,Zong:2006xj,Shi:2006dv}.  It is moot for QCD in-medium because no new divergences are encountered in a theory that is properly regularised in-vacuum.  In Sec.\,\ref{sec:chi} we therefore discuss the question of regularisation for the in-vacuum theory and present a number of model-independent results.  The scalar Ward identity plays an important role.  In order to illustrate the application of the chiral susceptibility we employ two simple models for the gap equation's kernel.  They are described in Sec.\,\ref{sec:models}.  The difference is expressed in the form of the quark-gluon vertex and in Sec.\,\ref{sec:results} we show that this difference has only a minor quantitative effect.  That both \emph{Ans\"atze} can be employed self-consistently owes again to the scalar Ward identity.  We wrap-up in Sec.\,\ref{last}.

\section{Chiral Susceptibility}
\label{sec:chi}
In QCD the vacuum quark condensate associated with a quark of flavour $f$ can be written \cite{Langfeld:2003ye}
\begin{equation}
\label{sigmaf}
\sigma_f(m_f;\zeta,\Lambda) = Z_4(\zeta,\Lambda) N_c {\rm tr}_{\rm D} \! \int_q^\Lambda\! S_f(q;m_f;\zeta)\,,
\end{equation}
where: $m_f(\zeta)$ is the renormalised current-quark mass, with $\zeta$ the renormalisation scale; $Z_4$ is the Lagrangian mass-term renormalisation constant, which depends implicitly on the gauge parameter; and $\int_q^\Lambda:= \int^\Lambda d^4 q/(2\pi)^4$ represents a suitable regularisation of the integral, with $\Lambda$ the regularisation mass-scale, which is taken to infinity as the last step in a complete calculation.

The dressed $f$-quark propagator, $S_f$ in Eq.\,(\ref{sigmaf}), is obtained from the gap equation\footnote{In our Euclidean metric:  $\{\gamma_\mu,\gamma_\nu\} = 2\delta_{\mu\nu}$; $\gamma_\mu^\dagger = \gamma_\mu$; $\gamma_5= \gamma_4\gamma_1\gamma_2\gamma_3$; $a \cdot b = \sum_{i=1}^4 a_i b_i$; and $P_\mu$ timelike $\Rightarrow$ $P^2<0$.}
\begin{equation}
S_f(p)^{-1} =  Z_2 \,(i\gamma\cdot p + m_f^{\rm bm}) + \Sigma_f(p)\,, \label{gendse} \end{equation}
with
\begin{equation}
\Sigma_f(p) = Z_1 \int^\Lambda_q\! g^2 D_{\mu\nu}(p-q) \frac{\lambda^a}{2}\gamma_\mu S_f(q) \frac{\lambda^a}{2}\Gamma^g_\nu(q,p) , \label{gensigma}
\end{equation}
where $D_{\mu\nu}(k)$ is the dressed-gluon propagator, $\Gamma^g_\nu(q,p)$ is the dressed-quark-gluon vertex, and $m_f^{\rm bm}$ is the $\Lambda$-dependent current-quark bare mass.  The quark-gluon-vertex and quark wave-function renormalisation constants, $Z_{1,2}(\zeta^2,\Lambda^2)$, also depend on the gauge parameter.  

The gap equation's solution has the form 
\begin{eqnarray} 
 S_f(p)^{-1} & = & i \gamma\cdot p \, A_f(p^2,\zeta^2) + B_f(p^2,\zeta^2) \,.
%
\label{sinvp} 
\end{eqnarray}
and the mass function $M_f(p^2)=B_f(p^2,\zeta^2)/A_f(p^2,\zeta^2)$ is renormalisation point independent.  The propagator is obtained from Eq.\,(\ref{gendse}) augmented by a renormalisation condition.  Since QCD is asymptotically free, the chiral limit is defined by 
\begin{equation}
Z_2(\zeta^2,\Lambda^2) \, m_f^{\rm bm}(\Lambda) \equiv 0\,,\; \forall \Lambda \gg \zeta\,,
\end{equation}
which is equivalent to requiring that the renormalisation point invariant current-quark mass is zero; i.e., $\hat m_f = 0$.  A mass-independent renormalisation scheme can then be implemented by fixing all renormalisation constants in the chiral limit \cite{Weinberg:1951ss}; namely, one requires  
\begin{equation}
\label{renormS} \left.S_f(p)^{-1}\right|_{p^2=\zeta^2} = i\gamma\cdot p \,.
\end{equation}
NB.\
\begin{equation}
Z_2(\zeta^2,\Lambda^2) \, m_f^{\rm bm}(\Lambda)=Z_4(\zeta^2,\Lambda^2) \, m_f(\zeta)\,.
\end{equation}

While it is not readily apparent from Eqs.\,(\ref{gendse}), (\ref{gensigma}), if one considers a $N_f$ flavour theory, 
then there is coupling between the gap equations for different flavoured quarks.  The coupling is driven by vertex corrections, as illustrated with a model in Ref.\,\cite{Cloet:2008fw}.  Absent this coupling it is impossible to obtain other than a mean field chiral symmetry restoring phase transition at nonzero temperature \cite{Holl:1998qs}.  We will return to this elsewhere but omit further substantial consideration herein.  Consequently, we usually omit the flavour label in all that follows.

The chiral susceptibility measures the response of a chiral order parameter to changes in current-quark mass.  Consider therefore
\begin{eqnarray}
\label{chif}
\chi(\zeta) & := & \left. \frac{\partial }{\partial m(\zeta)} \, \sigma(m;\zeta,\Lambda)\right|_{\hat m=0}\\
 &=& Z_4(\zeta,\Lambda) N_c {\rm tr}_{\rm D} \! \int_q^\Lambda\! \left.  \frac{\partial }{\partial m} S(q;m;\zeta)\right|_{\hat m=0}\,.
\end{eqnarray}
The order of integration and differentiation can be interchanged because the theory is properly regularised.  One can now use the Ward identity 
\begin{equation}
\frac{\partial }{\partial m} S(q;m;\zeta) = - S(q;m;\zeta) \Gamma_0 (q,0;\zeta) S(q;m;\zeta) \label{scalarWI}
\end{equation}
to obtain
\begin{equation}
\chi(\zeta) =   - Z_4(\zeta,\Lambda) N_c
%
{\rm tr}_{\rm D} \! \int_q^\Lambda\!
S(q;0;\zeta) \Gamma_0 (q,0;\zeta) S(q;0;\zeta)\,, \label{chim}
\end{equation}
wherein $\Gamma_0$ is the renormalised fully-dressed scalar vertex, which satisfies an inhomogeneous Bethe-Salpeter equation:
\begin{eqnarray}
\lefteqn{\Gamma_0(k,P;\zeta) = Z_4 \mbox{\boldmath $I_D$}} \nonumber\\
& + &
\int_q^\Lambda\! [S(q_+) \Gamma_0 (q,P) S(q_-)]_{sr} K_{tu}^{rs}(q,k;P)\,. \label{scalarBSE}
\end{eqnarray}
In this equation: $k$ is the relative and $P$ the total momentum of the quark-antiquark pair; $q_\pm = q \pm P/2$; $r,s,t,u$ represent colour and Dirac indices; and $K$ is the fully-amputated quark-antiquark scattering matrix.

In QCD the quantity $m^2 \omega_0(P)$ is a renormalisation point invariant, where $\omega_0(P)$ is the scalar vacuum polarisation.  In this product the term which multiplies $m(\zeta)^2$ is 
\begin{equation} 
\omega_0(P;\zeta) = Z_4 \, N_c{\rm tr}_{D} \int_q^\Lambda\! S(q_+) \Gamma_0(q,P) S(q_-)\,.
\end{equation}
Upon comparison with Eq.\,(\ref{chim}) one arrives at the general result:
\begin{equation}
\label{chimpol}
\chi(\zeta) = - \omega_0(P=0;\hat m=0,\zeta) \,.
\end{equation}

Hitherto we have not specified a regularisation procedure for the in-vacuum chiral susceptibility.  In this connection it is noteworthy that if a hard cutoff is used, then in the chiral limit of a noninteracting theory 
\begin{equation}
\label{Lambdadiverge}
- \omega_0^0(P=0;\hat m=0) = \frac{N_c}{4\pi^2} \Lambda^2.
\end{equation}
Following Ref.\,\cite{Langfeld:2003ye}, this result can be traced to the dependence on current-quark mass in Eq.\,(\ref{sigmaf}).  On the other hand, Pauli-Villars regularisation would yield zero as the result, and this is the procedure we recommend and employ in models that preserve the one-loop renormalisation-group behaviour of QCD.

One may implement a Pauli-Villars regularisation by introducing a pseudo-quark with large mass $m^{PQ}=\Lambda$, which is anticoupled to the scalar source \cite{Holl:2005vu}.  To be concrete, this means  
\begin{eqnarray} 
\nonumber\lefteqn{
\omega_0(P;\hat m=0,\zeta) = \lim_{\Lambda \to \infty} Z_4(\zeta,\Lambda) }\\
\nonumber
&\times & \, 
N_c{\rm tr}_{D} \int \frac{d^4 q}{(2\pi)^4} \bigg[
S(q_+;0) \Gamma_0(q,P;0) S(q_-;0)  \\
&& - S(q_+;\Lambda) \Gamma_0(q,P;\Lambda) S(q_-;\Lambda) \bigg].
\label{chiproper}
\end{eqnarray}
To proceed, one solves the gap equation, for the regular-quark and pseudo-quark, then uses the results to obtain the associated scalar vertices from Eq.\,(\ref{scalarBSE}), and finally computes $\chi(\zeta)$ from Eq.\,(\ref{chimpol}).
 
For the calculation of $\chi(\zeta)$ one only requires the scalar vertex at $P=0$, at which total momentum it has the general form
\begin{equation}
\label{G0vtx}
\Gamma_0(k,0;m) = i\gamma\cdot k \, C(k^2;m) + D(k^2;m)\,.
\end{equation}
Owing to Eq.\,(\ref{scalarWI}) it is not necessary in principle to solve the inhomogeneous Bethe-Salpeter equation because 
\begin{equation}
\label{CDAB}
C(k^2;m) = \frac{\partial}{\partial m} A(k^2;m)\,;\; 
D(k^2;m) = \frac{\partial}{\partial m} B(k^2;m)\,.
\end{equation}
Thus a solution of the gap equation suffices completely to fix $\Gamma_0(k,0)$.  The practical utility of this procedure is limited only by the time required to solve the gap equation numerically for a range of current-quark mass values and therefrom construct $C$, $D$. 

As another application of the scalar Ward identity, consider that in Landau gauge QCD one can write \cite{Lane:1974he,Politzer:1976tv}
\begin{eqnarray}
\lefteqn{ A(k^2;m) \stackrel{k^2\gg \Lambda_{\rm QCD}^2}{=}  1\,,}\\
\nonumber 
\lefteqn{B(k^2;m) \stackrel{k^2\gg \Lambda_{\rm QCD}^2}{=} 
\frac{\hat m}{(\frac{1}{2} \ln[k^2/\Lambda_{\rm QCD}^2])^{\gamma_m}}}\\
&-& \frac{2 \pi^2 \gamma_m}{3} \frac{\langle \bar q q \rangle^0}
{k^2 (\frac{1}{2} \ln[k^2/\Lambda_{\rm QCD}^2])^{1-\gamma_m}},
\end{eqnarray}
where $\gamma_m = 12/(33 - 2 N_f)$, with $N_f$ the number of active flavours, and $\langle \bar q q \rangle^0$ is the renormalisation-group-invariant chiral-limit vacuum quark condensate.  It therefore follows from Eq.\,(\ref{scalarWI}) that $C=0$ and 
\begin{eqnarray}
\nonumber\lefteqn{D(k^2;m) \stackrel{k^2\gg \Lambda_{\rm QCD}^2}{=} Z_4(\zeta^2,k^2)} \\
&+& \frac{2 \pi^2 \gamma_m}{3} \frac{X_\zeta^0}
{k^2 (\frac{1}{2} \ln[k^2/\Lambda_{\rm QCD}^2])^{1-\gamma_m}},
\end{eqnarray}
where 
\begin{equation}
X_\zeta^0 = -\frac{\partial \langle \bar q q \rangle^0}{\partial m(\zeta)}.
\label{Xqbq}
\end{equation}
Inserting Eqs.\,(\ref{G0vtx}) -- (\ref{Xqbq}) in Eq.\,(\ref{chiproper}) and using the fact that $3\gamma_m > 1$ for all $N_f$, then 
\begin{eqnarray}
\nonumber
\lefteqn{ -\omega_0(P=0;m=0,\zeta) = 
\lim_{\Lambda \to \infty} Z_4(\zeta,\Lambda) N_c } \\
&\times& \int_q^\Lambda\! \frac{2 \pi^2 \gamma_m}{3} \frac{X_\zeta^0}{q^2 (\frac{1}{2} \ln[q^2/\Lambda_{\rm QCD}^2])^{1-\gamma_m}},
\end{eqnarray}
where all terms in the integrand that integrate to a finite value have been dropped.  Thus
\begin{equation}
-\omega_0(P=0;m=0,\zeta) = X_\zeta^0 \ln[\zeta/\Lambda_{\rm QCD}])^{\gamma_m}
= -\frac{\partial \langle \bar q q \rangle_\zeta^0}{\partial m(\zeta)}.
\end{equation}
This brief analysis demonstrates the consistency of our definitions and procedure.  It also emphasises that the in-vacuum chiral susceptibility is a truly well-defined quantity.  Indeed, one sees that in principle its value is contained in the solution for the dressed scalar vertex.

\section{Gap equation models}
\label{sec:models}
A dialogue between DSE studies and results from numerical simulations of lattice-regularised QCD is providing important information about the kernel of QCD's gap equation; e.g., Refs.\,\cite{Bhagwat:2003vw,Alkofer:2003jj,Bhagwat:2004hn,Bhagwat:2004kjBhagwat:2006tu,%
Kizilersu:2006et,Kamleh:2007ud,Boucaud:2008ky,Cucchieri:2008fc}.  This body of work can be used to formulate reasonable \textit{Ans\"atze} for the dressed-gluon propagator and dressed-quark-gluon vertex in Eq.\,(\ref{gensigma}).  

In connection with such \emph{Ans\"atze} it is notable that the DSEs admit at least one nonperturbative symmetry-preserving truncation scheme \cite{Bhagwat:2004hn,Munczek:1994zz,Bender:1996bb}, which has enabled the proof of numerous exact results \cite{Maris:1997hd,Bicudo:2003fp,Holl:2004fr,Holl:2005vu,McNeile:2006qy,Bhagwat:2007ha,%
Ivanov:1997yg,Ivanov:1998ms,Bhagwat:2006xi,Chang:2008sp}.  It also provides a starting point for the formulation of reliable models that can be used to illustrate those results and make predictions with readily quantifiable errors \cite{Roberts:2007jh,Maris:1997tm,Maris:1999nt,Bloch:2002eq,Maris:2002mt,Eichmann:2008ae,Eichmann:2008ef}.  
The \emph{Ans\"atze} are typically implemented by writing
\begin{eqnarray}
\nonumber \lefteqn{Z_1 g^2 D_{\rho \sigma}(p-q) \Gamma_\sigma^a(q,p)} \\
& =& {\cal G}((p-q)^2) \, D_{\rho\sigma}^{\rm free}(p-q) \frac{\lambda^a}{2}\Gamma_\sigma(q,p)\,, \label{KernelAnsatz}
\end{eqnarray}
wherein $D_{\rho \sigma}^{\rm free}(\ell)$ is the Landau-gauge free gauge-boson propagator, ${\cal G}(\ell^2)$ is a model effective-interaction and $\Gamma_\sigma(q,p)$ is a vertex \textit{Ansatz}.  

In one widely used approach to in-vacuum physics, ${\cal G}(\ell^2)$ is chosen such that the one-loop renormalisation group behaviour of QCD is preserved and the vertex is written
\begin{equation}
\label{rainbowV}
\Gamma_\sigma(q,p) = \gamma_\sigma\,.
\end{equation}
This is the basis for a renormalisation-group-improved rainbow-ladder (RL) truncation of QCD's DSEs  \cite{Maris:1997tm,Maris:1999nt,Bloch:2002eq,Maris:2002mt,%
Eichmann:2008ae,Eichmann:2008ef}.  

One can alternatively employ \emph{Ans\"atze} for the vertex whose diagrammatic content is unknown.  A class of such models that has hitherto seen much use can be characterised by \cite{Ball:1980ay}
\begin{eqnarray}
\label{bcvtx}
\nonumber \lefteqn{i\Gamma_\sigma(k,\ell)  =
i\Sigma_A(k^2,\ell^2)\,\gamma_\sigma + (k+\ell)_\sigma }\\
&\times &
\left[\frac{i}{2}\gamma\cdot (k+\ell) \,
\Delta_A(k^2,\ell^2) + \Delta_B(k^2,\ell^2)\right] \!,
\end{eqnarray}
where 
\begin{eqnarray}
\Sigma_F(k^2,\ell^2)& = &\frac{1}{2}\,[F(k^2)+F(\ell^2)]\,,\;\\
\Delta_F(k^2,\ell^2) &=&
\frac{F(k^2)-F(\ell^2)}{k^2-\ell^2}\,,
\label{DeltaF}
\end{eqnarray}
with $F= A, B$; viz., the scalar functions in Eq.\,(\ref{sinvp}).  This \emph{Ansatz} satisfies the vector Ward-Takahashi identity and is often referred to as the BC vertex.
 
One has a Slavnov-Taylor identity for the quark-gluon vertex in QCD, not a Ward-Takahashi identity.  Hence, Eq.\,(\ref{bcvtx}) is not necessarily an improvement over Eq.\,(\ref{rainbowV}).  A comparison between results obtained with the different \emph{Ans\"atze} is nevertheless useful in identifying those outcomes which might be robust.  

Herein we employ a simplified form of the renormalisation-group-improved effective interaction in Refs.\,\cite{Maris:1997tm,Maris:1999nt,Bloch:2002eq,Maris:2002mt,%
Eichmann:2008ae,Eichmann:2008ef}; viz., we retain only that piece which expresses the long-range behaviour ($s=k^2$):
\begin{equation}
\label{IRGs}
\frac{{\cal G}(s)}{s} = \frac{4\pi^2}{\omega^6} \, D\, s\, {\rm e}^{-s/\omega^2}.
\end{equation}
This is a finite width representation of the form introduced in Ref.\,\cite{mn83}, which has been rendered as an integrable regularisation of $1/k^4$ \cite{mm97}.  Equation~(\ref{IRGs}) delivers an ultraviolet finite model gap equation.  Hence, the regularisation mass-scale can be removed to infinity and the renormalisation constants set equal to one.  

\begin{table}[t]
\caption{Results obtained for selected quantities with $\omega=0.5\,$GeV, and the vertex and $D$ parameter value indicated:
$A(0)$, $M(0)$ are $p=0$ in-vacuum values of the scalar functions defined in connection with Eq.\,(\ref{sinvp}); the vacuum quark condensate is defined with $m=0$ in Eq.\,(\ref{sigmaf}); and $\chi$ is obtained from Eq.\,(\protect\ref{chipractical}).
$A(0)$ is dimensionless but all other entries are quoted in GeV.
The calculations reported herein were performed in the chiral limit.
\label{Table:Para1} 
}
\begin{center}
\begin{tabular*}
{\hsize}
{|l@{\extracolsep{0ptplus1fil}}
|c@{\extracolsep{0ptplus1fil}}
|c@{\extracolsep{0ptplus1fil}}
|c@{\extracolsep{0ptplus1fil}}
|c@{\extracolsep{0ptplus1fil}}
|c@{\extracolsep{0ptplus1fil}}|} \hline
\rule{0em}{3ex} 
Vertex & $\sqrt D$ & $A(0)$ & $M(0)$ & $-(\langle \bar q q \rangle^0)^{1/3}$ & $\sqrt\chi$   \\\hline
Eq.\,(\ref{rainbowV}), {\rm RL} & 1 & 1.3 & 0.40 & 0.25 & 0.39  \\\hline
Eq.\,(\ref{bcvtx}), {\rm BC} & $\frac{1}{\surd 2}$ & 1.1 & 0.28 & 0.26 & 0.28 \\\hline
\end{tabular*}
\end{center}
\end{table}

The active parameters in Eq.\,(\ref{IRGs}) are $D$ and $\omega$ but they are not independent.  In reconsidering a renormalisation-group-improved rainbow-ladder fit to a selection of ground state observables \cite{Maris:1999nt}, Ref.\,\cite{Maris:2002mt} noted that a change in $D$ is compensated by an alteration of $\omega$.  This feature has further been elucidated and exploited in Refs.\,\cite{Cloet:2008fw,Eichmann:2008ae,Eichmann:2008ef}.  For $\omega\in[0.3,0.5]\,$GeV, with the interaction specified by Eqs.\,(\ref{KernelAnsatz}), (\ref{rainbowV}) and (\ref{IRGs}), fitted in-vacuum low-energy observables are approximately constant along the trajectory
\begin{equation}
\label{gluonmass}
\omega D  = (0.8 \, {\rm GeV})^3 =: m_g^3\,.
\end{equation}
Herein, we employ $\omega=0.5\,$GeV.  Therefore $D=m_g^3/\omega=1.0\,$GeV$^2$ corresponds to what might be called the real-world reference value for Eq.\,(\ref{rainbowV}).

It is impossible at present to explore whether the behaviour expressed in connection with Eq.\,(\ref{gluonmass}) is also realised with the interaction specified by Eqs.\,(\ref{KernelAnsatz}), (\ref{bcvtx}) and (\ref{IRGs}) because the BC vertex, Eq.\,(\ref{bcvtx}), cannot yet be used in the implementation of a symmetry preserving DSE truncation.  We note, however, that $D=0.5\,$GeV$^2$ reproduces the physical value of the in-vacuum condensate, last row of Table~\ref{Table:Para1}.  We therefore identify this as the real-world reference value for Eq.\,(\ref{bcvtx}).

In association with Eqs.\,(\ref{CDAB}) we noted that it might be time consuming to solve the gap equation numerically for a range of values of current-quark mass and therefrom construct $C$, $D$.  It is therefore useful if that procedure can be avoided.  In a class of cases, it can.  

From the gap equation
\begin{equation}
\label{Gamma0practical}
\Gamma_0(p,0) = \frac{\partial}{\partial m} S^{-1}(p) = \mbox{\boldmath $I$}_{\rm D} + \frac{\partial}{\partial m} \Sigma(p)\,.
\end{equation}
One can always write
\begin{equation}
\frac{\partial}{\partial m} \Sigma(p) = \Gamma_0^1(p,0) + \Gamma_0^2(p,0)\,,
\end{equation}
where, using Eq.\,(\ref{scalarWI}), 
\begin{eqnarray}
\lefteqn{\Gamma_0^1(p,0)  = - \int_q g^2 D_{\mu\nu}(p-q)} \nonumber \\
&  \times & \frac{\lambda^a}{2}\gamma_\mu S(q) \Gamma_0(q,0) S(q) \frac{\lambda^a}{2}\Gamma_\nu(q,p)\,,\\
\lefteqn{\Gamma_0^2(p,0)  = \int_q g^2 D_{\mu\nu}(p-q)} \nonumber \\
&  \times & \frac{\lambda^a}{2}\gamma_\mu S(q) \frac{\lambda^a}{2} \frac{\partial}{\partial m} \Gamma_\nu(q,p)\,.
\end{eqnarray}
Now, with any \emph{Ansatz} for the dressed-quark-gluon vertex whose current-quark-mass-dependence is completely determined by that of the dressed-quark propagator, $\frac{\partial}{\partial m} \Gamma_\nu(q,p)$ has a closed form involving $C$, $D$.  In these cases Eq.\,(\ref{Gamma0practical}) is a self-contained pair of coupled integral equations for $C$, $D$.  NB.\ Eqs.\,(\ref{rainbowV}), (\ref{bcvtx}) are both in this class.  In the former case $\Gamma_0^2(p,0) \equiv 0$, while in the latter
\begin{eqnarray}
\nonumber \lefteqn{\frac{\partial}{\partial m} \Gamma_\nu(q,p) = 
i\Sigma_C(q^2,p^2)\,\gamma_\nu + (q+p)_\nu }\\
&\times &
\left[\frac{i}{2}\gamma\cdot (q+p) \,
\Delta_C(q^2,p^2) + \Delta_D(q^2,p^2)\right] \!.
\end{eqnarray}

As noted above, the interaction in Eq.\,(\ref{IRGs}) does not preserve the one-loop behaviour of QCD but, indeed, delivers an ultraviolet finite gap equation.  Nevertheless, Eq.\,(\ref{Lambdadiverge}) demonstrates that a regularisation procedure would be required for the vacuum polarisation even if one were dealing with a free field theory.  For QCD, in which Schwinger functions carry anomalous dimensions, the Pauli-Villars scheme works well.  However, that procedure, as defined above, fails when the vacuum polarisation is evaluated with Schwinger functions generated by Eq.\,(\ref{IRGs}).  A simple alternative can be constructed.

The vector vacuum polarisation is given by
\begin{equation}
\omega^V_{\mu \nu}(P) = N_c {\rm tr}_{\rm D} \! \int_q^\Lambda\!
i \gamma_\mu S(q_+) i\Gamma_\nu (q,P) S(q_-)\,,
\end{equation}
where $\Gamma_\nu$ is the vector quark-antiquark vertex; i.e., the vector analogue of $\Gamma_0$, which satisfies an inhomogeneous Bethe-Salpeter equation similar to Eq.\,(\ref{scalarBSE}).\footnote{Recall that with Eq.\,(\protect\ref{IRGs}) the regularisation mass-scale should be removed to infinity and the renormalisation constants set equal to one.} This polarisation couples to the photon and hence must be transverse and contain no mass term.  Consider therefore  \cite{Chang:2008sp}
\begin{eqnarray}
\lefteqn{\frac{1}{4}\omega^V_{\mu \mu}(P=0) = \frac{N_c}{4} {\rm tr}_{\rm D} \! \int_q^\Lambda\! i \gamma_\mu S(q) i\Gamma_\mu (q,0) S(q)} \nonumber \\
&=& - \frac{N_c}{4} {\rm tr}_{\rm D} \! \int_q^\Lambda\! 
i \gamma_\mu  \frac{\partial}{\partial q_\mu} S(q) \rule{3em}{0ex}\nonumber \\
&=& - 2 N_c \int_q^\Lambda\! \frac{1}{q^2} \frac{d}{dq^2}\left[ (q^2)^2 \sigma_V(q^2) \right],\rule{3em}{0ex}
\label{vectorvacuum}
\end{eqnarray}
where the vector Ward identity was used in the second line and we have written the dressed quark propagator in the form
\begin{equation}
S(p) = -i \gamma\cdot p \, \sigma_V(p^2) + \sigma_S(p^2)\,.
\end{equation}
Evaluating Eq.\,(\ref{vectorvacuum}) with a hard cutoff in a noninteracting theory one obtains
\begin{equation}
- \frac{1}{4}\omega^{V 0}_{\mu \mu}(P=0) = \frac{N_c}{8 \pi^2} \Lambda^2;
\end{equation}
i.e., precisely the same sort of divergence as encountered in the scalar vacuum polarisation.  Naturally, any regularisation scheme that preserves the vector Ward-Takahashi would yield zero as the result \cite{Burden:1991uh}.

In connection with Eq.\,(\ref{IRGs}) we therefore define the regularised in-vacuum chiral susceptibility as 
\begin{eqnarray}
\label{chipractical} 
\lefteqn{\chi = -\omega_0(0;m=0,\Lambda) + \frac{1}{2}\omega^{V}_{\mu \mu}(0;m=0,\Lambda)} \\
\nonumber & = & - \frac{N_c}{4\pi^2} \int_0^\infty ds \, s \,\bigg\{
D(s) [\sigma_s(s)^2 - s \sigma_s(s)^2] \\
&& + 2 s \, C(s) \sigma_V(s) \sigma_S(s)  + 2 \sigma_V(s) +  s \sigma_V^\prime(s) \bigg\}
\label{chipracticalR} 
\end{eqnarray}
It will be recognised that so long as each term in Eq.\,(\ref{chipractical}) is independently regularised in a valid fashion, which one can safely assume, then this is a nugatory transformation.  After all, we have only subtracted zero: the photon is massless and hence in any valid scheme it must be that $\omega^{V}_{\mu \mu}(0;0,\Lambda)=0$.  On the other hand, with this definition one can proceed to calculate the susceptibility without it being necessary to actually specify a regularisation scheme.  (NB.\ The last two contributions in the integrand act as the Pauli-Villars terms.  $Z_4=1$ in the ultraviolet-finite model.)  

\begin{figure}[t]
\vspace*{-5ex}

\centerline{\includegraphics[clip,width=0.5\textwidth]{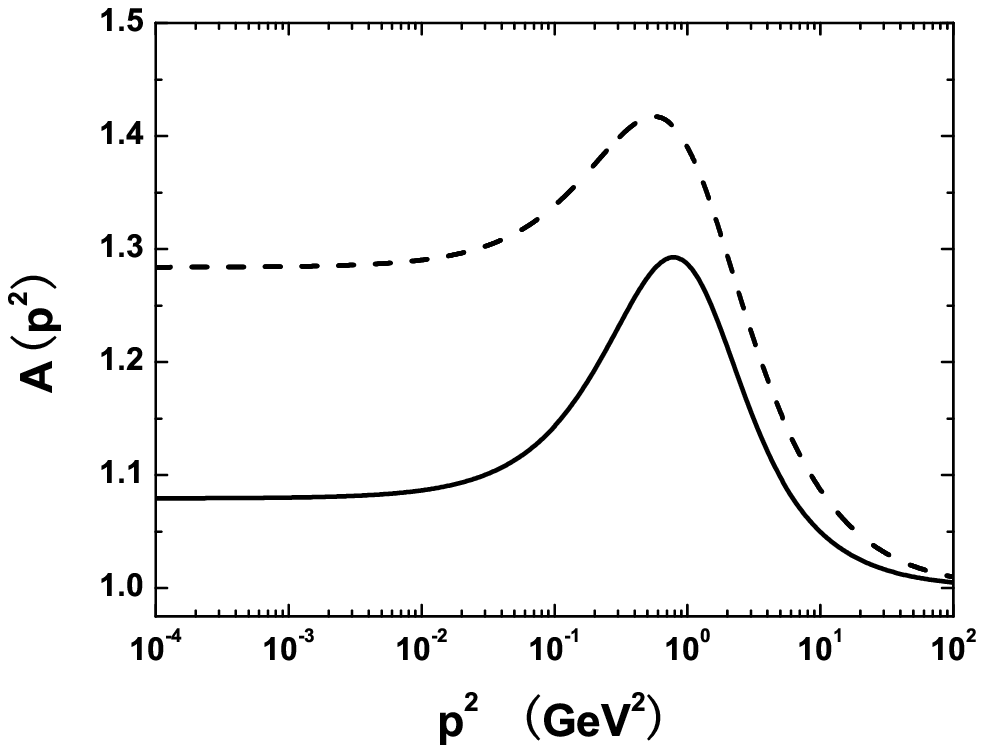}}
\vspace*{-7ex}

\centerline{\includegraphics[clip,width=0.5\textwidth]{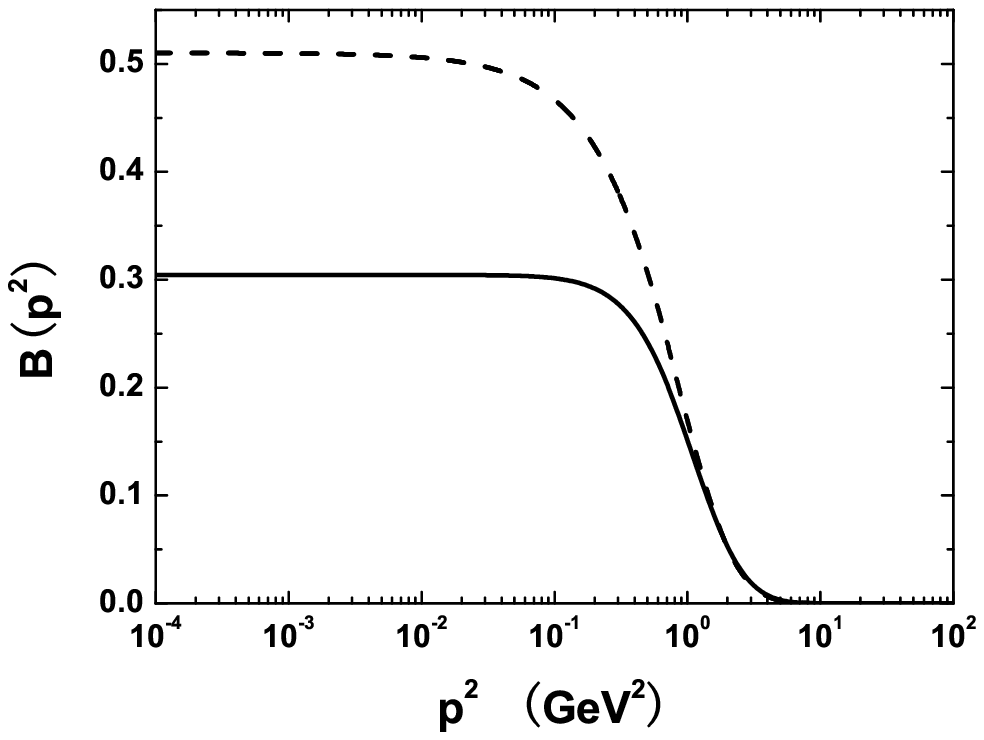}}
\vspace*{-5ex}

\caption{\label{aamm} Dressed quark propagator.  \emph{Upper panel} -- $A(p^2) = 1/Z(p^2)$, where $Z$ is the wave function renormalisation function.  
\emph{Lower panel} -- $B(p^2)$, the scalar piece of the dressed-quark self-energy.  
In both panels, \underline{Dashed curve}: calculated in rainbow-ladder truncation, Eq.\,(\protect\ref{rainbowV}), with ${\cal I}=D/\omega^2=4$; \underline{solid curve}: calculated with BC vertex \emph{Ansatz}, Eq.\,(\protect\ref{bcvtx}), and ${\cal I}=2$.}
\end{figure}

\begin{figure}[t]
\vspace*{-5ex}

\includegraphics[clip,width=0.5\textwidth]{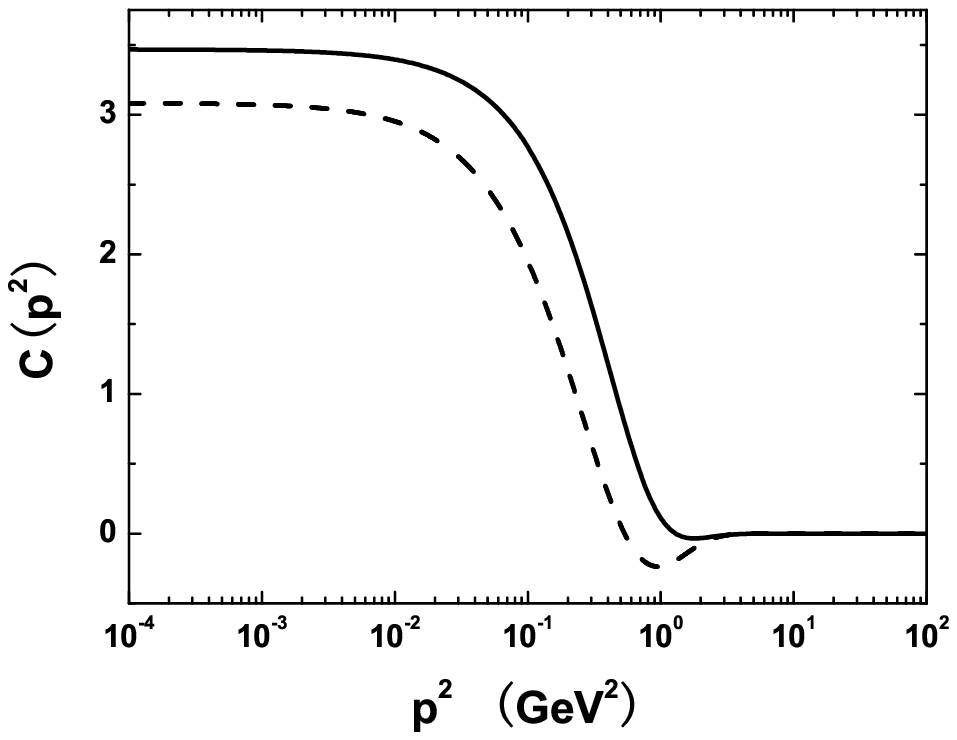}
\vspace*{-9ex}

\includegraphics[clip,width=0.5\textwidth]{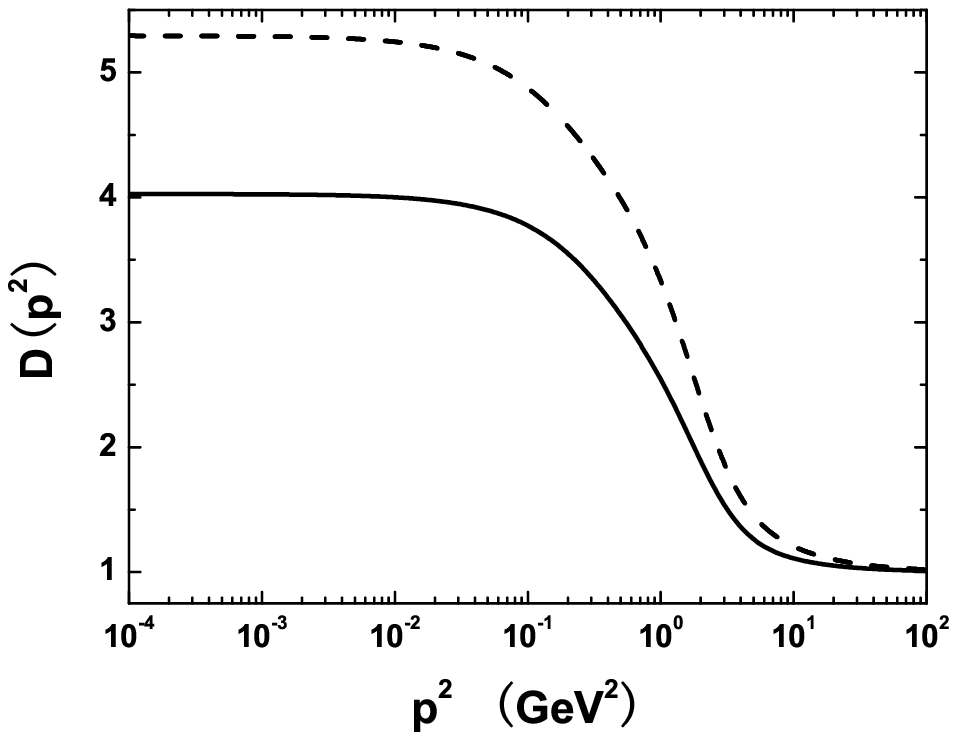}
\vspace*{-5ex}

\caption{\label{ccdd} 
$P=0$ scalar vertex, Eq.\,(\protect\ref{G0vtx}: \emph{upper panel} -- $C(p^2)$, 
\emph{lower panel} -- $D(p^2)$.
In both panels, \underline{Dashed curve}: calculated in rainbow-ladder truncation, Eq.\,(\protect\ref{rainbowV}), with ${\cal I}=D/\omega^2=4$; \underline{solid curve}: calculated with BC vertex \emph{Ansatz}, Eq.\,(\protect\ref{bcvtx}), and ${\cal I}=2$.}
\end{figure}

\section{Illustration}
\label{sec:results}
It is now a straightforward matter to solve for the dressed-quark propagator and the $P=0$ scalar vertex.  Owing to the Gau{\ss}ian form of Eq.\,(\ref{IRGs}), all relevant integrations converge rapidly.  In Fig.\,\ref{aamm} we plot the functions obtained through solving the gap equation and in Fig.\,\ref{ccdd} those which describe the $P=0$ scalar vertex.  The results were obtained with the appropriate real-world value of the interaction strength: $D=1\,$GeV$^2$ with Eq.\,(\ref{rainbowV}) and $D=0.5\,$GeV$^2$ with Eq.\,(\ref{bcvtx}).  

It is apparent in Fig.\,\ref{aamm} that the vertex \emph{Ansatz} has a quantitative impact on the magnitude and pointwise evolution of the gap equation's solution.  That this should be anticipated is plain from Ref.\,\cite{Burden:1991gd}.  Moreover, the pattern of behaviour can be understood from Ref.\,\cite{Bhagwat:2004hn}: the feedback arising through the $\Delta_B$ term in the BC vertex, Eq.\,(\ref{bcvtx}), absent in Eq.\,(\ref{rainbowV}), always acts to alter the domain upon which $A(p^2)$ and $M(p^2)$ differ significantly in magnitude from their respective free-particle values.  Since $C(p^2)$ and $D(p^2)$ are derived quantities, their behaviour does not require explanation.

\begin{figure}[t]
\vspace*{-5ex}

\begin{center}
\includegraphics[clip,width=0.5\textwidth]{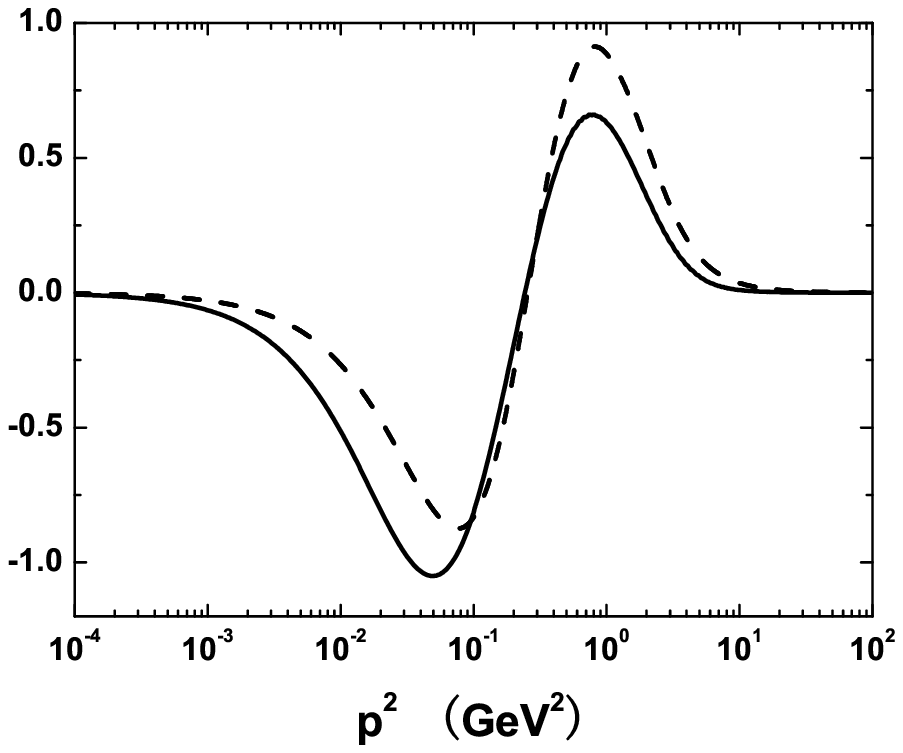}
\vspace*{-7ex}

\caption{\label{integrandRW} Integrand in Eq.\,(\protect\ref{chipracticalR}) --  
\underline{Dashed curve}: calculated in rainbow-ladder truncation, Eq.\,(\protect\ref{rainbowV}), with ${\cal I}=D/\omega^2=4$; \underline{solid curve}: calculated with BC vertex \emph{Ansatz}, Eq.\,(\protect\ref{bcvtx}), and ${\cal I}=2$.}
\end{center}
\end{figure}

The integrand in Eq.\,(\ref{chipracticalR}) is depicted in Fig.\,\ref{integrandRW} for each vertex \emph{Ansatz} at the associated real-world interaction strength.  The resulting chiral susceptibilities are presented in Table\,\ref{Table:Para1}.  Two things are immediately apparent in the figure.  First, the regularisation has served to eliminate the far-ultraviolet tail of the integrand, thereby ensuring convergence of the integral.  We have varied the detailed form of the regularising subtraction; e.g., using free-field propagators and vertices instead of gap and Bethe-Salpeter equation solutions.  The pointwise behaviour of the integrand is little altered.
Second, at real-world values of the interaction strength for both \emph{Ans\"atze} the integrand has negative support in the infrared and positive support for $p^2 \gtrsim (0.5\,{\rm GeV})^2$.  We will subsequently return to this point.

\begin{figure}[t]
\vspace*{-5ex}

\includegraphics[clip,width=0.5\textwidth]{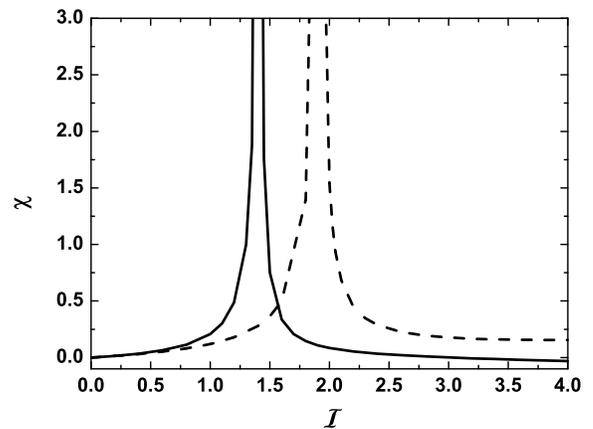}
\vspace{-1cm} \caption{\label{chiI} Dependence of the chiral susceptibility on the interaction strength in Eq.\,(\protect\ref{IRGs}); viz., ${\cal I}:= D/\omega^2$: \emph{dashed curve}, RL vertex, Eq.\,(\protect\ref{rainbowV}); \emph{solid curve}, BC vertex, Eq.\,(\protect\ref{bcvtx}).}
\end{figure}

In Fig.\,\ref{chiI} we depict the evolution of the chiral susceptibility with increasing interaction strength, ${\cal I}=D/\omega^2$.  The behaviour may readily be understood.  
For ${\cal I}=0$ one has a noninteracting theory and the ``vacuum'' is unperturbed by a small change in current-quark mass.  Hence, the susceptibility is zero.
The susceptibility grows with increasing ${\cal I}$ because the interaction is attractive in the $\bar q q$ channel and therefore magnifies the associated pairing.  This is equivalent to stating that the scalar vertex is enhanced above its free field value.

The growth continues and accelerates until, at some critical value, ${\cal I}={\cal I}_c$, the susceptibility becomes infinite.  Those critical values are:
\begin{equation}
\label{critIc}
\begin{array}{l|c|c} 
    & {\rm Eq.}\,(\ref{rainbowV}),\; {\rm RL} & {\rm Eq.}\,(\ref{bcvtx}),\; {\rm BC} \\
{\cal I}_c & 1.93 & 1.41
\end{array}\,.
\end{equation}
This divergence is easily understood.  The models we've defined contain a dimensionless parameter, ${\cal I}$, which characterises the interaction strength, and a current-quark mass, which is an explicit source of chiral symmetry breaking.  In the general theory of phase transitions the latter is analogous to an external magnetic field while $1/{\cal I}$ is kindred to a temperature.  

Consider the free energy for such theories
$f(t,m)\,,$
where $t = [1-{\cal I}_c/{\cal I}]$.  If such a theory possesses a second-order phase transition, then the free energy is an homogeneous function of its arguments in the neighbourhood of $t = 0 = m$.  From this it follows that the theory's magnetisation exhibits the following behaviour (e.g., see the appendix of Ref.\,\protect\cite{Blaschke:1998mp}):
\begin{equation}
M(t,0)=t^\beta,\; t\to 0^+,
\end{equation}
and the associated magnetic susceptibility evolves according to
\begin{equation}
M(0,m)\propto m^{-(1-1/\delta)}.
\end{equation}
NB.\ For a mean-field theory $\beta = 1/2$ and $\delta = 3$.  

The nature of the critical interaction strength is now plain.  
In the class of theories we're considering, the vacuum quark condensate is analogous to the magnetisation.  It is attended by the chiral susceptibility.\footnote{The dressed-quark mass function evaluated at $p^2=0$ and the pion's leptonic decay constant are equivalent order parameters.}  
For ${\cal I}<{\cal I}_c$ the interaction has insufficient strength to generate a nonzero scalar term in the dressed-quark self-energy in the absence of a current-quark mass; namely, dynamical chiral symmetry breaking (DCSB) is impossible and the model realises chiral symmetry in the Wigner-Weyl mode.
That changes at ${\cal I}_c$, so that for ${\cal I}>{\cal I}_c$ a $B\neq 0$ solution is always possible.  Moreover, the behaviour of the susceptibility shows that each model undergoes a second-order phase transition and realises chiral symmetry in the Nambu-Goldstone mode for interaction strengths above their respective values of ${\cal I}_c$. 

These observations emphasise the chiral susceptibility's usefulness.  With the vertex of Eq.\,(\ref{rainbowV}) one can explicitly construct the pressure\footnote{The pressure is defined as the negative of the effective-action.  A system's ground state is that configuration for which the pressure is a global maximum or, equivalently, the effective-action is a global minimum.  These statements are elucidated in Ref.\,\protect\cite{haymaker}.} and therewith show that for ${\cal I}>{\cal I}_c$ the DCSB solution is dynamically favoured because it corresponds to the configuration of maximum pressure.  On the other hand, the diagrammatic content of the vertex in Eq.\,(\ref{bcvtx}) is unknowable and hence one cannot derive an expression for the pressure.  In this case one may rely on the behaviour of the susceptibility to conclude that DCSB is favoured, as illustrated in Ref.\,\cite{Zhao:2008zz} within the Nambu--Jona-Lasinio model.

Once past the critical point, the susceptibility decreases as ${\cal I}$ increases.  This is because the magnitude of the condensate order-parameter grows in tandem with ${\cal I}$ and therefore the influence of any perturbation associated with a current-quark mass must steadily diminish.\footnote{On the other hand, if at fixed coupling one were to increase the current-quark mass from zero, the dynamical component of the order parameter would steadily be suppressed in response \protect\cite{Holl:2005st,Chang:2006bm}.}

As in the earlier figures, Fig.\,\ref{chiI} displays results with a quantitative dependence on the vertex \emph{Ansatz}.  This is easily understood \cite{Bhagwat:2004hn}: at any given, common interaction strength in Eq.\,(\ref{IRGs}), the BC vertex \emph{Ansatz}, Eq.\,(\ref{bcvtx}), amplifies attraction in the $\bar q q$ channel in comparison with that obtained in the rainbow truncation, Eq.\,(\ref{rainbowV}).  

\begin{figure}[t]
\vspace*{-5ex}

\includegraphics[clip,width=0.5\textwidth]{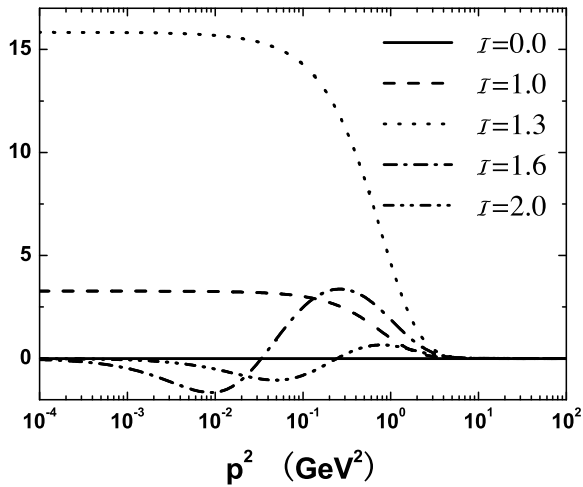}
\vspace{-1cm} \caption{\label{integrandD} Influence of ${\cal I}:= D/\omega^2$ on the pointwise behaviour of the dimensionless integrand in Eq.\,(\protect\ref{chipracticalR}), evaluated with the BC vertex, Eq.\,(\protect\ref{bcvtx}).}
\end{figure}

In Fig.\,(\ref{integrandD}) we depict the evolution with ${\cal I}$ of the integrand in Eq.\,(\ref{chipractical}).  This illustrates the manner by which the susceptibility's behaviour is generated.  For ${\cal I}<{\cal I}_c$ the integrand is a nonnegative monotonically decreasing function whose pointwise magnitude grows rapidly as ${\cal I}\to {\cal I}_c^-$.  Plainly, given the scalar Ward identity, Eq.\,(\ref{scalarWI}), the functions characterising the scalar vertex are the origin of this rapid growth: the vertex exhibits a singularity at ${\cal I}_c$.  
For ${\cal I}> {\cal I}_c$ the integrand possesses a zero.  Its origin is the existence on this domain of a nonzero mass function in the chiral limit, which provides for a negative contribution owing to the $D\,\sigma_S^2$ term in Eq.\,(\ref{chipracticalR}).

The behaviour of the scalar vertex and the susceptibility that we have described contains additional, valuable information.  At ${\cal I}=0$ one has a free field theory and no bound state in the scalar channel, nor any other.  As ${\cal I}$ increases away from zero, attraction appears in the scalar channel and with it a $\bar q q$ correlation.  Nevertheless, for ${\cal I}<{\cal I}_c$ in the chiral limit the dressed-quarks remain massless.  Hence, absent confinement, zero is the only realisable mass for a bound state in the scalar channel.  However, since $\chi$ involves the scalar vertex evaluated at $P=0$ and $0<\chi<\infty$, it is apparent that no scalar bound state exists when ${\cal I} < {\cal I}_c$.  

On the other hand, the chiral limit susceptibility exhibits a singularity at ${\cal I} = {\cal I}_c$.  This feature is tied to a pole at $P^2=0$ that appears in the scalar vertex.  Thus at the critical interaction strength a massless bound state appears in the scalar channel.  Moreover, since at precisely ${\cal I}_c$ the dressed-quark mass is still zero, then a degenerate; i.e., massless, pseudoscalar bound state appears simultaneously.  It is evident in the evolution of $\chi$ that the scalar bound state becomes steadily more massive as ${\cal I}$ increases away from the critical value.

In terms of $t = [1-{\cal I}_c/{\cal I}]$ the pattern of behaviour just described is similar in many respects to that exhibited by such models at nonzero temperature; e.g., Ref.\,\cite{Maris:2000ig}.  The key difference is the absence of anything with the character of a screening mass for ${\cal I}<{\cal I}_c$.  The existence of a screening mass allows a pole in the scalar vertex for temperatures above critical.

\section{Summary}
\label{last}
We demonstrated that the in-vacuum chiral susceptibility is given by the scalar vacuum polarisation at zero total momentum.  The scalar Ward identity is essential in deriving this result from the expression for the vacuum quark condensate.  
One can rigorously define the in-vacuum susceptibility by employing a Pauli-Villars regularisation procedure.  While this provides the most practical means by which to calculate the susceptibility, the derivation shows that, in principle, all information about the susceptibility is contained within the $P=0$ behaviour of the dressed scalar vertex.  This confirms \emph{a posteriori} that a valid regularisation of the scalar vacuum polarisation is always possible.  

One must solve the gap equation in order to calculate the chiral susceptibility.  The kernel of that equation involves the dressed quark-gluon vertex.  If an \emph{Ansatz} for that vertex is employed whose diagrammatic content is unknown, then a consistent Bethe-Salpeter kernel cannot currently be constructed.  In general this prevents the calculation of all vertices and in particular those that couple to colour singlet $\bar q$-$q$ channels.  However, owing to the Ward identities, that problem is circumvented at $P=0$ in the scalar and vector vertices.  This is a material point.  For example, herein it means that no matter what form is assumed for the dressed-quark-gluon vertex and, indeed, for the kernel of the gap equation, the consistent chiral susceptibility can always be constructed.  
Additional simplifications follow if one employs an \emph{Ansatz} for the quark-gluon vertex whose current-quark-mass-dependence is completely determined by that of the dressed-quark propagator.  These features enabled us to provide a consistent calculation of the chiral susceptibility for two rather different vertex \emph{Ans\"atze}.  The results were consistent and the minor quantitative differences easily understood.  

A key result is that the susceptibility can be used to demarcate the domain of coupling strength within a theory upon which chiral symmetry is dynamically broken.  For couplings below the associated critical value and in the absence of confinement, there are no bound states.  Degenerate massless scalar and pseudoscalar bound-states appear at the critical coupling.  The behaviour of the susceptibility shows that the scalar bound-state becomes massive as the coupling increases above this value whereas, naturally, the Goldstone mode remains massless in the chiral limit.  

We described and used a simple model for the gluon propagator in the gap equation and hence in computing the susceptibility.  Nevertheless, no qualitative feature of our results will change should a form be used that, for example, expresses properly the one-loop renormalisation group behaviour of QCD.  That is demonstrated by calculations which are now underway.  

In addition, effects that might arise through flavour mixing in the gap equation are quashed by the vertex \emph{Ans\"atze} we considered.  Such mixing is generated by nontrivial correlations in the vertex.  It is certainly important in-medium and should also be considered in-vacuum, where it is likely to alter the behaviour of the susceptibility in the neighbourhood of the critical coupling strength.


\begin{acknowledgments}
This work was supported by: the National Natural Science Foundation of China, under Contract Nos.\ 10425521, 10675007, 10705002 and 10775069; the Major State Basic Research Development Program, under Contract No.\ G2007CB815000; the Key Grant Project of the Chinese Ministry of Education, under contact No. 305001; the Department of Energy, Office of Nuclear Physics, contract no.\ DE-AC02-06CH11357; and the Gordon Godfrey Fund of the School of Physics at the University of New South Wales.
\end{acknowledgments}

\end{document}